\documentclass[aps,prd,preprint,showpacs,superscriptaddress]{revtex4}
\usepackage{graphics}
\usepackage{psfig}
\usepackage{tabularx}

\begin{document}
\title{Systematic study of the $SU(3)_c\otimes SU(4)_L\otimes U(1)_X$ gauge symmetry}
\author{William A. Ponce} 
\affiliation{Instituto de F\'\i sica, Universidad de Antioquia,
A.A. 1226, Medell\'\i n, Colombia}
\author{Luis A. S\'anchez}
\affiliation{Escuela de F\'\i sica, Universidad Nacional de Colombia,
A.A. 3840, Medell\'\i n, Colombia}


\begin{abstract}
We carry a systematic study of possible extensions of the standard model based on the gauge group $SU(3)_c\otimes SU(4)_L\otimes U(1)_X$. We consider both models with particles with exotic electric charges and models which do not contain exotic electric charges neither in the gauge boson sector nor in the fermion sector. For the first case an infinite number of models can, in principle, be constructed, while the restriction to non-exotic electric charges only allows for eight different anomaly-free models. Four of them are three-family models in the sense that anomalies cancel by an interplay between the three families, and another two are one-family models where anomalies cancel family by family as in the standard model. The remaining two are two-family models.
\end{abstract}

\maketitle

\section{Introduction}
The standard model (SM), based on the gauge group $SU(3)_c\otimes SU(2)_L\otimes U(1)_Y$, with $SU(2)_L\otimes U(1)_Y$ hidden and $SU(3)_c$
confined, can be extended in several different ways: first, by adding new fermion fields (adding a right-handed neutrino field constitute its simplest extension and has profound consequences, as the implementation of the see-saw mechanism, and the enlarging of the possible number of local abelian symmetries that can be gauged simultaneously); second, by augmenting the scalar sector to more than one Higgs representation, and third by enlarging the local gauge
group. In this last direction $SU(4)_L\otimes U(1)_X$ as a flavor group
has been considered  in the literature \cite{su4a,su4b,cota,FaRi1,FaRi2,little,kong,pgs,spp} which, among its best features, provides with an alternative to the problem of the number $N_f$ of fermion families in Nature, in the sense that anomaly cancellation is achieved when $N_f=N_c=3, \; N_c$ being the number of colors of $SU(3)_c$ \cite{su4b}. Moreover, this gauge structure has been used recently in order to implement the little Higgs mechanism \cite{little,kong}.

In this work a systematic study of the $SU(3)_c\otimes SU(4)_L\otimes U(1)_X$ local gauge symmetry (hereafter the 3-4-1 theory) shows that, by restricting the fermion field representations to particles without exotic electric charges and by paying due attention to anomaly cancellation, a few
different models are obtained, while by relaxing the condition of
nonexistence of exotic electric charges, an infinite number of models can
be generated, all of them with particles with exotic electric charges (quarks with electric charges $5/3$ and $-4/3$, for example), as the ones considered in Refs.~\cite{su4a,su4b,cota,FaRi1,FaRi2}.

An analysis along the same lines has previosly been carried out for the extension of the SM based on the gauge group $SU(3)_c\otimes SU(3)_L\otimes U(1)_X$ \cite{331} (the so-called 3-3-1 models), with the result that there exit only ten different anomaly free models with particles with only ordinary electric charges. Two of them are one family models and have been partially studied in Ref.~\cite{spm}. Eight are models for three families, and the supersymmetric version of one of them has been considered in Ref.~\cite{331susy}, while some interesting phenomenology of these models has been studied in Ref.~\cite{sher}. Well motivated 3-3-1 models which include particles with exotic electric charges have also been considered in the literature \cite{su3L}.

This paper is organized as follows. In Sec.~\ref{sec2} we introduce the characteristics of the gauge group; in Sec.~\ref{sec3} we present the eight different anomaly-free models without exotic electric charges that can be constructed and study their scalar and gauge boson sectors; in Sec.~\ref{sec4} we discuss models with exotic electric charges, and in the last section we present our conclusions. 

\section{\label{sec2}The Model}
We assume that the electroweak group is 
$SU(4)_L\otimes U(1)_X\supset SU(3)_L \otimes U(1)_Z 
\supset SU(2)_L \otimes U(1)_Y$, where the gauge structure 
$SU(3)_L\otimes U(1)_Z$ refers to the one presented in Ref.~\cite{331}. 
We also assume that the left handed quarks (color triplets), left-handed 
leptons (color singlets) and scalars, transform either under the 4 or 
the $\bar{4}$ fundamental representations of $SU(4)_L$. As in the SM, $SU(3)_c$ is vectorlike.

In $SU(4)_L\otimes U(1)_X$, the most general expression for the electric charge generator is a linear combination of the four diagonal
generators of the gauge group
\begin{equation}\label{ch}
Q=aT_{3L}+\frac{1}{\sqrt{3}}bT_{8L}+ \frac{1}{\sqrt{6}}cT_{15L}+ XI_4, 
\end{equation} 
where $T_{iL}=\lambda_{iL}/2$, being $\lambda_{iL}$ the Gell-Mann matrices
for $SU(4)_L$ normalized as  Tr$(\lambda_i\lambda_j)=2\delta_{ij}$,
$I_4=Dg(1,1,1,1)$ is the diagonal $4\times 4$ unit matrix, and $a$, $b$ 
and $c$ are free parameters to be fixed next. Notice that we can absorb  
an eventual coefficient for $X$ in its definition.

If we assume that the usual isospin $SU(2)_L$ of the SM is such that
$SU(2)_L\subset SU(4)_L$, and we demand for accomodating each family of SM fermions into different fundamental representations $4$ or $\bar{4}$ of $SU(4)_L$, then $a=1$ and we have just a two-parameter set
of models, all of them characterized by the values of $b$ and $c$. So, 
Eq.~(\ref{ch}) allows for an infinite number of models in the context of the 3-4-1 theory, each one associated to particular values of the parameters $b$ and $c$, with characteristic signatures that make them different from each other. 

There are a total of 24 gauge bosons in the gauge group under
consideration, 15 of them associated with $SU(4)_L$ which can be written as
\begin{equation}
\frac{1}{2}\lambda_\alpha A^\alpha_\mu =\frac{1}{\sqrt{2}} 
\left(\begin{array}{cccc}D^0_{1\mu} & W^+_\mu & K^{(b+1)/2}_\mu & X^{(3+b
+2c)/6}_\mu\\ 
W^-_\mu & D^0_{2\mu} &  K^{(b-1)/2}_{1\mu} &  V^{(-3+b+2c)/6}_\mu \\
K^{-(b+1)/2}_\mu & K^{-(b-1)/2}_{1\mu} & D^0_{3\mu} & Y^{-(b-c)/3}_\mu\\
X^{-(3+b+2c)/6}_\mu & V^{(3-b-2c)/6}_\mu & Y^{(b-c)/3}_\mu & 
D^0_{4\mu} \end{array}\right),  
\label{gauge}
\end{equation}
where $D^{0\mu}_1=A_3^\mu/\sqrt{2}+A_8^\mu/\sqrt{6}+A_{15}^\mu/\sqrt{12}$;
$D^{0\mu}_2=-A_3^\mu/\sqrt{2}+A_8^\mu/\sqrt{6}+A_{15}^\mu/\sqrt{12}$;
$D^{0\mu}_3=-2A_8^\mu/\sqrt{6}+A_{15}^\mu/\sqrt{12}$, and 
$D^{0\mu}_4=-3 A_{15}^\mu/\sqrt{12}$. The upper indices in the gauge bosons in the former expression stand for the electric charge of the corresponding 
particle, some of them functions of the $b$ and $c$ parameters as they 
should be. 

Different from the SM where only the abelian $U(1)_Y$ factor is
anomalous, in the 3-4-1 theory both, $SU(4)_L$ and $U(1)_X$ are anomalous
($SU(3)_c$ is vectorlike). So, special combinations of multiplets must be
used in each particular model in order to cancel the possible
anomalies, and obtain renormalizable models. The triangle anomalies
we must take care of are: $[SU(4)_L]^3$, $[SU(3)_c]^2U(1)_X$, 
$[SU(4)_L]^2U(1)_X$, $[grav]^2U(1)_X$ and $[U(1)_X]^3$.

Let us now see how the charge operator in Eq.~(\ref{ch}) acts on the
representations 4 and $\bar{4}$ of $SU(4)_L$:

\begin{eqnarray}\nonumber
Q[4]
&=&Dg.\left(\frac{1}{2}+\frac{b}{6}+\frac{c}{12}+X, -\frac{1}{2}+\frac{b}{6}+\frac{c}{12}+X, -\frac{2b}{6}+\frac{c}{12}+X, -\frac{3c}{12}+X\right),\\ \label{eq3}
Q[\bar{4}]
&=&Dg.\left(-\frac{1}{2}-\frac{b}{6}-\frac{c}{12}+X, \frac{1}{2}-\frac{b}{6}-\frac{c}{12}+X, \frac{2b}{6}-\frac{c}{12}+X, \frac{3c}{12}+X\right).
\end{eqnarray}

\section{\label{sec3}Models Without Exotic Electric Charges}
Notice, from Eq.~(\ref{eq3}), that if we demand for gauge bosons with electric charges $0, \pm 1$ only, there are not more than four different possibilities for the simultaneous values of $b$ and $c$; they are: 
$b= c = 1$; $b = c = -1$; $b = 1$, $c = -2$, and $b = -1$, $c = 2$.
 
It is clear that, if we accommodate the known
left-handed quark and lepton isodoublets in the two upper components of 4
and $\bar{4}$ (or $\bar{4}$ and 4), do not allow for electrically charged
antiparticles in the two lower components of the multiplets (antiquarks
violate $SU(3)_c$, and $e^+, \mu^+$ and $\tau^+$ violate lepton number at
tree level) and forbid the presence of exotic electric charges in the
possible models, then the electric charge of the third and fourth
components in $4$ and $\bar{4}$ must be equal either to the charge of the
first and/or second component, which in turn implies that $b$ and $c$ can
take only the four sets of values stated above. So, these four sets of
values for $b$ and $c$ are necessary and sufficient conditions in order to
exclude exotic electric charges in the fermion sector too.

A further analysis also shows that models with $b=c=-1$ are equivalent,
via charge conjugation, to models with $b=c=1$. Similarly, models 
with $b=-1,\; c=2$ are equivalent to models with $b=1,\; c=-2$. So, 
with the constraints impossed, we have only two different sets of models; 
those for $b=c=1$ and those for $b=1, \; c=-2$. 

3-4-1 models without exotic electric charges of both classes have been proposed in Refs.~\cite{little,kong} as viable models to implement the little Higgs mechanism.

\subsection{Models for $b=c=1$}

Let us start defining the following complete sets of spin 1/2 Weyl spinors 
(complete in the sense that each set contains its own charged 
antiparticles):
\begin{itemize}
\item $S_1^q=\{(u,d,D,D^\prime)_L\sim [3,4,-\frac{1}{12}], \; 
u_L^c \sim [\bar{3}, 1, -\frac{2}{3}], \; 
d_L^c \sim [\bar{3},1,\frac{1}{3}], \; 
D_L^c \sim [\bar{3}, 1, \frac{1}{3}], \; 
D_L^{\prime c} \sim [\bar{3}, 1, \frac{1}{3}]\}.$ 
\item $S_2^q=\{(d,u,U,U^\prime)_L\sim [3,\bar{4},\frac{5}{12}], \; 
u_L^c \sim [\bar{3}, 1, -\frac{2}{3}], \; 
d_L^c \sim [\bar{3},1,\frac{1}{3}], \; 
U_L^c \sim [\bar{3}, 1, -\frac{2}{3}], \; 
U_L^{\prime c} \sim [\bar{3},1,-\frac{2}{3}]\}.$ 
\item $S_3^l=\{ (\nu^0_e, e^-, E^-, E^{\prime -})_L \sim 
[1,4,-\frac{3}{4}],\;
e^+_L \sim [1,1,1], \; E^+_L \sim [1,1,1], \; E^{\prime +}_L \sim 
[1,1,1]\}.$ 
\item $S_4^l = \{ (E^+, N^0_1, N^0_2, N^0_3)_L\sim 
[1, 4, \frac{1}{4}], \; E^-_L\sim [1,1,-1]\}.$ 
\item $S_5^l = \{ (e^-, \nu^0_e, N^0, N^{\prime 0})_L\sim 
[1, \bar{4}, -\frac{1}{4}], \; e^+_L\sim [1,1,1]\}.$ 
\item $S_6^l=\{(N^0, E_1^+, E_2^+, E_3^+)_L \sim [1, \bar{4}, 
\frac{3}{4}],\;
E_{1L}^- \sim [1,1,-1], \; E_{2L}^- \sim [1,1,-1], \; E_{3L}^- \sim 
[1,1,-1]\}.$
\end{itemize}

Taking into account that each set includes charged particles together with
their corresponding antiparticles, and since $SU(3)_c$ is vectorlike, the
anomalies $[grav]^2U(1)_X, \; [SU(3)_c]^3$ and $[SU(3)_c]^2U(1)_X$
automatically vanish. So, we only have to take care of the remaining three
anomalies whose values are shown in Table~\ref{tab1}. From this table several anomaly free models can be constructed. Let us see.
 
There are two three family structures which are:
\begin{itemize}
\item Model {\bf A} = $2S_1^q \oplus S_2^q \oplus 3S_5^l$. (This model 
has been analyzed in Ref.~\cite{pgs}). 
\item Model {\bf B} = $S_1^q \oplus 2S_2^q \oplus 3S_3^l$. 
\end{itemize}

We find only one two family structure given by:
Model {\bf C} = $S_1^q \oplus S_2^q \oplus S_3^l \oplus S_5^l$. 

A one family model can not be directly extracted from $S_i, \; i=1,2,...,6$, but we can check that the following particular arrangement is an anomaly free one family structure:
Model {\bf D}=$S_1^q\oplus (e^-,\nu_e^0, N^0, N^{\prime 0})_L \oplus 
(E_1^-, N_1^0, N_2^0, N_3^0)_L \oplus (N_4^0, E_1^+, e^+, E_2^+)_L \oplus E_2^-.$ As it can be checked, this model reduces to the model in 
Ref.~\cite{spm} for the breaking chain $SU(4)_L\otimes U(1)_X\longrightarrow SU(3)_L\otimes U(1)_\alpha\otimes U(1)_X\longrightarrow SU(3)_L\otimes U(1)_Z$, for the value $\alpha = 1/12$. In an analogous way, other one family models with more exotic charged leptons can also be constructed.

\begin{table*}
\caption{\label{tab1}Anomalies for models with $b$=$c$=1}
\begin{ruledtabular}
\begin{tabular}{lcccccc} 
Anomaly & $S_1^q$ & $S_2^q$ & $S_3^l$ & $S_4^l$ & $S_5^l$ & $S_6^l$ \\ 
\hline
$[U(1)_X]^3$ & $-9/16$ & $-27/16$ & 21/16 & $-15/16$ & 15/16 & $-21/16$ \\
$[SU(4)_L]^2U(1)_X$ & $-1/4$ & 5/4 & $-3/4$ & 1/4 & $-1/4$ & 3/4 \\
$[SU(4)_L]^3 $ & 3 & $-3$ & 1 & 1 & $-1$ & $-1$ \\  
\end{tabular}
\end{ruledtabular}
\end{table*}

\subsection{Models for $b=1, c=-2$}
As in the previous case, let us define the following complete sets of spin 
1/2 Weyl spinors:
\begin{itemize}
\item $S_1^{\prime q}=\{(u,d,D,U)_L\sim [3,4,\frac{1}{6}], \; 
u_L^c \sim [\bar{3}, 1, -\frac{2}{3}], \; 
d_L^c \sim [\bar{3},1,\frac{1}{3}], \; 
D_L^c \sim [\bar{3}, 1, \frac{1}{3}], \; 
U_L^{c} \sim [\bar{3}, 1, -\frac{2}{3}]\}.$ 
\item $S_2^{\prime q}=\{(d,u,U,D)_L\sim [3,\bar{4},\frac{1}{6}], \; 
u_L^c \sim [\bar{3}, 1, -\frac{2}{3}], \; 
d_L^c \sim [\bar{3},1,\frac{1}{3}], \; 
U_L^c \sim [\bar{3}, 1, -\frac{2}{3}], \; 
D_L^{c} \sim [\bar{3},1,\frac{1}{3}]\}.$ 
\item $S_3^{\prime l}=\{ (\nu^0_e, e^-, E^-, N^0)_L \sim 
[1,4,-\frac{1}{2}],\;
e^+_L \sim [1,1,1], \; E^+_L \sim [1,1,1]\}.$ 
\item $S_4^{\prime l} = \{ (e^-, \nu^0_e, N^0, E^-)_L\sim 
[1, \bar{4}, -\frac{1}{2}], \; e^+_L\sim [1,1,1],\; E^+_L\sim [1,1,1]\}.$ 
\item $S_5^{\prime l} = \{(E^+, N^0_1, N^0_2, e^+)_L\sim 
[1, 4, \frac{1}{2}], \; E^-_L\sim [1,1,-1],\; e^-_L\sim [1,1,-1]\}.$ 
\item $S_6^{\prime l}=\{ (N_3^0, E^+, e^+, N_4^0)_L \sim 
[1, \bar{4}, \frac{1}{2}],\;
E_L^- \sim [1,1,-1], \; e_{L}^- \sim [1,1,-1]\}.$
\end{itemize}

For these sets the anomalies $[grav]^2U(1)_X, \; [SU(3)_c]^3$ and
$[SU(3)_c]^2U(1)_X$ vanish.  The other anomalies are shown in 
Table~\ref{tab2}. Again, several anomaly free models can be constructed from this table. Let us see.

\begin{table*}
\caption{\label{tab2}Anomalies for models with $b=1, \; c=-2$}
\begin{ruledtabular}
\begin{tabular}{lcccccc} 
Anomaly & $S_1^q$ & $S_2^q$ & $S_3^l$ & $S_4^l$ & $S_5^l$ & $S_6^l$ \\ 
\hline
$[U(1)_X]^3$ & $-3/2$ & $-3/2$ & 3/2 & $ 3/2 $ & -3/2 & $-3/2 $ \\
$[SU(4)_L]^2U(1)_X$ & 1/2 & 1/2 & $-1/2$ & $-1/2$ & 1/2 & 1/2 \\
$[SU(4)_L]^3 $ & 3 & $-3$ & 1 & $-1$ & 1 & $-1$  \\ 
\end{tabular}
\end{ruledtabular}
\end{table*}

We find two three family structures which are:
\begin{itemize}
\item Model {\bf E} = $2S_1^{\prime q} \oplus S_2^{\prime q} \oplus 
3S_4^{\prime l}$. (This model has been studied in Ref.~\cite{spp}). 
\item Model {\bf F} = $S_1^{\prime q} \oplus 2S_2^{\prime q} \oplus 
3S_3^{\prime l}$. 
\end{itemize}

We again find only one two family structure given by:
Model {\bf G} = $S_1^{\prime q} \oplus S_2^{\prime q}\oplus 
S_3^{\prime l}\oplus S_4^{\prime l}$. 

Two one family models can be constructed using $S^\prime_i, \; i=1,....,6$. 
They are:
\begin{itemize}
\item Model {\bf H}= $S^{\prime q}_2\oplus 2S^{\prime l}_3\oplus S^{\prime l}_5$.
\item Model {\bf I}= $S^{\prime q}_1\oplus 2S^{\prime l}_4\oplus S^{\prime l}_6$.
\end{itemize}

\subsection{The Scalar and Gauge Boson Sectors}

\subsubsection{Models for $b=c=1$}

If our aim is to break the symmetry following the pattern
$SU(3)_c\otimes SU(4)_L\otimes U(1)_X  \rightarrow  SU(3)_c\otimes SU(3)_L\otimes U(1)_X \rightarrow SU(3)_c\otimes SU(2)_L\otimes U(1)_Y  
\rightarrow SU(3)_c\otimes U(1)_Q$,
we must introduce, at least, the following three Higgs scalars \cite{pgs}: 
$\phi_1[1,4,-3/4]$ with a vacuum expectation value (VEV) aligned in the 
direction $\langle\phi_1\rangle=(v,0,0,0)^T$; $\phi_2[1,\bar{4},-1/4]$ 
with a VEV aligned as $\langle\phi_2\rangle=(0,0,V,0)^T$ and 
$\phi_3[1,\bar{4},-1/4]$
with a VEV aligned as $\langle\phi_3\rangle=(0,0,0,V^\prime)^T$, with the
hierarchy $V\sim V^\prime >> v\sim 174$ GeV (the electroweak breaking 
scale).

Now, the $SU(4)_L$ gauge boson sector for $b=c=1$, as can be seen from Eq.~(\ref{gauge}), is given by
\[\frac{1}{2}\lambda_\alpha A^\alpha_\mu=\frac{1}{\sqrt{2}}\left(
\begin{array}{cccc}D^0_{1\mu} & W^{+}_\mu & K^{+}_\mu & X^{+}_\mu\\ 
W^{-}_\mu & D^{0}_{2\mu} &  K^{0}_{1\mu} &  V^{0}_\mu\\
K^{-}_\mu & K^{\prime 0}_{1\mu} & D^{0}_{3\mu} & Y^{0}_\mu\\
X^{-}_\mu & V^{\prime 0}_\mu & Y^{\prime 0}_\mu & D^{0}_{4\mu} \end{array}\right).\]

After breaking the symmetry with $\langle\phi_1\rangle +
\langle\phi_2\rangle+ \langle\phi_3\rangle$ and using for the covariant
derivative for 4-plets $iD^\mu=i\partial^\mu-g \lambda_\alpha
A^\mu_\alpha/2-g'XB^\mu$, where $g$ and $g^\prime$ are the $SU(4)_L$ and 
$U(1)_X$ gauge coupling constants, respectively, we get the following mass 
terms for the charged gauge bosons: 
$M^2_{W^\pm}=g^2v^2/2$, $M^2_{K^\pm}=g^2(v^2+V^2)/2$, 
$M^2_{X^\pm}=g^2(v^2+V^{\prime 2})/2$,
$M^2_{K^0_1(K^{\prime 0}_1)}=g^2V^2/2$, $M^2_{V^0(V^{\prime0})}=g^2V^{\prime 2}/2$ and $M^2_{Y^0(Y^{\prime0})}=g^2(V^2+V^{\prime 2})/2$. 
Since $W^\pm$ does not mix with $K^\pm$ or with $X^\pm$ we have that 
$v\approx 174$ GeV as in the SM.

For the four neutral gauge bosons we get mass terms of the form 
\begin{eqnarray}\nonumber
M&=&\frac{g^2}{2}\Big\{V^2 \left(\frac{g'B^\mu}{2g}
-\frac{2A_8^\mu}{\sqrt{3}}+\frac{A^\mu_{15}}{\sqrt{6}}\right)^2
+ V^{\prime 2}\left(\frac{g'B^\mu}{2g}-\frac{3 
A^\mu_{15}}{\sqrt{6}}\right)^2\\ 
& & +v^2\left(A^\mu_3
+\frac{A_8^\mu}{\sqrt{3}}+\frac{A^\mu_{15}}{\sqrt{6}}
-\frac{3g'B^\mu}{2g}\right)^2\Big\}.
\end{eqnarray}
\noindent 
$M$ is a $4 \times 4$ matrix with one zero eigenvalue
corresponding to the photon. Once the photon field has been identified, we
remain with a $3 \times 3$ mass matrix for three neutral gauge bosons
$Z^\mu$, $Z^{'\mu}$ and $Z^{''\mu}$. For the particular case $V^\prime=V$,
the field $Z^{\prime \prime \mu}= A_8^\mu / \sqrt{3}-
\sqrt{2/3}A_{15}^\mu$ decouples from the other two and acquires a  
squared mass $(g^2/2)V^2$. By diagonalizing the remaining $2 \times 2$ mass matrix we get other two physical neutral gauge bosons which are defined through the mixing angle $\theta$ between $Z_\mu$ and $Z'_\mu$: 
\begin{equation}\nonumber
Z_1^\mu=Z_\mu \cos\theta+Z'_\mu \sin\theta, \qquad
Z_2^\mu=-Z_\mu \sin\theta+Z'_\mu \cos\theta, 
\end{equation} 
where
\begin{equation} \label{tan} \tan(2\theta) = - \frac{2 \sqrt{2} C_W}
{\sqrt{1+2 \delta^2}\left[1+ \frac{2V^2}{v^2}C_W^4- \frac{2}{1+2
\delta^2}C^2_W \right]}, \end{equation}
with $\delta= g'/(2g)$.

\noindent  
The photon field $A^\mu$ and the fields $Z_\mu$ and $Z'_\mu$ are given by
\begin{eqnarray} \nonumber
A^\mu&=&S_W A_3^\mu
+ C_W\left[\frac{T_W}{\sqrt{3}}\left(A_8^\mu+
\frac{A_{15}^\mu}{\sqrt{2}}\right)+(1-T_W^2/2)^{1/2}B^\mu\right]\; , \nonumber \\  
Z^\mu&=& C_W A_3^\mu - S_W\left[\frac{T_W}{\sqrt{3}}\left(A_8^\mu+
\frac{A_{15}^\mu}{\sqrt{2}}\right)+(1-T_W^2/2)^{1/2}B^\mu\right] \; , \nonumber \\ \label{fzzp}
Z^{\prime \mu}&=&\sqrt{\frac{2}{3}}(1-T_W^2/2)^{1/2}\left(A_8^\mu+
\frac{A_{15}^\mu}{\sqrt{2}}\right)-\frac{T_W}{\sqrt{2}}B^\mu.
\end{eqnarray}
\noindent
$S_W=2 \delta /\sqrt{6 \delta^2 + 1}$ and $C_W$ are the sine and
cosine of the electroweak mixing angle respectively, and
$T_W=S_W/C_W$. We can also identify the $Y$ hypercharge associated
with the SM abelian gauge boson as
\begin{equation}\label{y}
Y^\mu=\frac{T_W}{\sqrt{3}}\left(A_8^\mu+
\frac{A_{15}^\mu}{\sqrt{2}}\right)+(1-T_W^2/2)^{1/2}B^\mu.
\end{equation}

\subsubsection{Models for $b=1, c=-2$}

In this case, in order both to break the symmetry following the scheme $SU(3)_c\otimes SU(4)_L\otimes U(1)_X  \rightarrow  SU(3)_c\otimes SU(3)_L\otimes U(1)_X \rightarrow SU(3)_c\otimes SU(2)_L\otimes U(1)_Y  
\rightarrow SU(3)_c\otimes U(1)_Q$ and, at the same time, to give mass to the fermion fields (a problem that, in any case, is model dependent), the following four Higgs scalars must be introduced \cite{spp}: 
$\phi_1[1,\bar{4},-1/2]$ with a Vacuum Expectation Value (VEV) aligned in the direction $\langle\phi_1\rangle=(0,v,0,0)^T$; $\phi_2[1,\bar{4},-1/2]$ 
with a VEV aligned as $\langle\phi_2\rangle=(0,0,V,0)^T$;
$\phi_3[1,4,-1/2]$ with a VEV aligned in the direction $\langle\phi_3\rangle=(v^\prime,0,0,0)^T$, and $\phi_4[1,4,-1/2]$ with a VEV aligned as $\langle\phi_4\rangle=(0,0,0,V^\prime)^T,$ with the
hierarchy $V\sim V^\prime >> \sqrt{v^2 + v^{\prime 2}} \simeq 174$~GeV.

For $b=1$ and $c=-2$, the 15 gauge fields associated with $SU(4)_L$ can be written as (see Eq.~(\ref{gauge}))
\[\frac{1}{2}\lambda_\alpha A^\alpha_\mu=\frac{1}{\sqrt{2}}\left(
\begin{array}{cccc}D^{0}_{1\mu} & W^{+}_\mu & K^{+}_\mu & X^{0}_\mu\\ 
W^{-}_\mu & D^{0}_{2\mu} &  K^{0}_{1\mu} &  V^{-}_\mu\\
K^{-}_\mu & K^{\prime 0}_{1\mu} & D^{0}_{3\mu} & Y^{-}_\mu\\
X^{\prime 0}_\mu & V^{+}_\mu & Y^{+}_\mu & D^{0}_{4\mu} \end{array}\right). \]

After breaking the symmetry with $\langle\phi_1\rangle +
\langle\phi_2\rangle+ \langle\phi_3\rangle+ \langle\phi_4\rangle$, we get the following mass terms for the charged gauge bosons: 
$M^2_{W^\pm}=g^2(v^2+v^{\prime 2})/2$, 
$M^2_{K^\pm}=g^2(v^{\prime 2}+V^2)/2$, 
$M^2_{V^\pm}=g^2(v^2+V^{\prime 2})/2$,
$M^2_{Y^\pm}=g^2(V^2+V^{\prime 2})/2$,
$M^2_{K^0_1(K^{\prime 0}_1)}=g^2(v^2 + V^2)/2$, 
and $M^2_{X^0(X^{\prime 0})}=g^2(v^{\prime 2}+V^{\prime 2})/2$. 
Since $W^\pm$ does not mix with the other charged bosons we have that 
$\sqrt{v^2+v^{\prime 2}}\approx 174$ GeV.

For the four neutral gauge bosons we now get mass terms of the form 
\begin{eqnarray}
M&=&\frac{g^2}{2}\Big\{V^2 \left(\frac{g'B^\mu}{g}
-\frac{2A_8^\mu}{\sqrt{3}}+\frac{A^\mu_{15}}{\sqrt{6}}\right)^2 + V^{\prime 2}\left(\frac{g'B^\mu}{g}+\frac{3 
A^\mu_{15}}{\sqrt{6}}\right)^2 \nonumber \\
& & +v^{\prime 2}\left(A^\mu_3
+\frac{A_8^\mu}{\sqrt{3}}+\frac{A^\mu_{15}}{\sqrt{6}}
-\frac{g'B^\mu}{g}\right)^2 + v^2 \left(\frac{g'B^\mu}{g}-A^\mu_3
+\frac{A_8^\mu}{\sqrt{3}}+\frac{A^\mu_{15}}{\sqrt{6}}\right)^2 \Big\},
\end{eqnarray}
\noindent 
which is a $4 \times 4$ matrix with one zero eigenvalue
corresponding to the photon. After extracting the photon, a $3 \times 3$ mass matrix remains for three neutral gauge bosons
$Z^\mu$, $Z^{'\mu}$ and $Z^{''\mu}$. Let us choose $V^\prime=V$ and $v^\prime = v$ in order to simplify matters. For this particular case the field $Z^{''\mu}= 2A_8^\mu /\sqrt{6}+A_{15}^\mu/\sqrt{3}$ decouples from the other two and acquires a squared mass $(g^2/2)(V^2+v^2)$. By diagonalizing the remaining $2 \times 2$ mass matrix we get other two physical neutral gauge bosons which are defined through the mixing angle $\theta$ between $Z_\mu$ and $Z'_\mu$, which is now given by 
\begin{equation} \label{tan2} 
\tan(2\theta) =  \frac{S_W^2 \sqrt{C_{2W}}}
{(1+S_W^2)^2 + \frac{V^2}{v^2}C_W^4 - 2},
\end{equation}
\noindent 
where $S_W=g^\prime/\sqrt{2g^{\prime 2}+g^2}$ is the sine of the electroweak mixing angle.

The photon field $A^\mu$ and the fields $Z_\mu$ and $Z'_\mu$ are then given by
\begin{eqnarray} \nonumber
A^\mu&=&S_W A_3^\mu + C_W\left[\frac{T_W}{\sqrt{3}}\left(A_8^\mu-
2\frac{A_{15}^\mu}{\sqrt{2}}\right)+(1-T_W^2)^{1/2}B^\mu\right]\;,\nonumber \\  
Z^\mu&=& C_W A_3^\mu - S_W\left[\frac{T_W}{\sqrt{3}}\left(A_8^\mu-
2\frac{A_{15}^\mu}{\sqrt{2}}\right)+(1-T_W^2)^{1/2}B^\mu\right] \; , \nonumber \\ \label{fzzp4}
Z^{'\mu}&=&\frac{1}{\sqrt{3}}(1-T_W^2)^{1/2}\left(A_8^\mu-
2\frac{A_{15}^\mu}{\sqrt{2}}\right)-T_W B^\mu,
\end{eqnarray}
\noindent
and the $Y$ hypercharge associated with the SM abelian gauge boson is
\begin{equation}\label{y2}
Y^\mu=\frac{T_W}{\sqrt{3}}\left(A_8^\mu-
2\frac{A_{15}^\mu}{\sqrt{2}}\right)+(1-T_W^2)^{1/2}B^\mu.
\end{equation}

\section{\label{sec4}Models With Exotic Electric Charges}
We now relax the condition of non-existence of exotic electric charges in the 3-4-1 extension of the SM. As mentioned in Sec.~\ref{sec2}, the parameters $b$ and $c$ in Eq.~(\ref{ch}) are now arbitrary and we then have, in principle, an infinite number of possible embeddings of the SM gauge group into $SU(3)_c\otimes SU(4)_L\otimes U(1)_X$. A particular embedding depends on the physical motivations of the 3-4-1 model to be constructed.

It is well known that the enlargement of the symmetry group of the SM usually leads to fermions in large multiplets having fractionary electric charges different from $\pm 2/3$ and $\pm 1/3$ for exotic quarks and integer electric charges different from $0$ and $\pm 1$ for exotic leptons, and to new gauge bosons with electric charges larger than $1$ and/or fractionary (leptoquarks). Several 3-4-1 models have been constructed in the literature \cite{su4a,su4b,cota,FaRi1,FaRi2} which contain exotic electric charges only in the quark sector, while leptons have ordinary electric charges and gauge bosons have integer electric charges. We will restrict ourselves to this type of models and, for the sake of comparison with the models in the previous section, we will sketch the model in Ref.~\cite{su4b}.

We start by noticing from Eq.~(\ref{gauge}) that if we demand for gauge bosons with integer electric charges, then a first condition is that the $b$ parameter must be different from zero and that the allowed values for it are the (positive or negative) odd intergers. Now, the 3-4-1 model in 
Ref.~\cite{su4b} is constructed with the goal of including right-handed neutrinos in the theory in such a way that each family of leptons belongs to the same multiplet in the representation $4$ of $SU(4)_L$ and includes the charge conjugate fields of a SM $SU(2)_L$ doublet; so, the correct embedding of $SU(2)_L$ doublets implies to start with $(\nu_a, e_a, \nu^c_{a}, e^c_{a})_L\sim [1,4,X_{aL}]$ (where $c$ stands for charge conjugation, and $a=1,2,3$ is a family index). This condition and Eq.~(\ref{ch}) provide three simultaneous equations from which the values of $b$, $c$ and of the hypercharge $X_{aL}$ of these three multiplets can be obtained. The results are: $b=-1$, $c=-4$, and $X_{aL}=0$.

With $b$ and $c$ fixed we can again resort to Eq.~(\ref{ch}) in order to see which the quark content of a $4$-plet and a $\bar{4}$-plet in the model must be. For a correct embedding of $SU(2)_L$ doublets, the first and second members in a $4$-plet must have electric charges $2/3$ and $-1/3$, respectively, while in a $\bar{4}$-plet these charges must be $-1/3$ and $2/3$. Again, these condition and Eq.~(\ref{ch}) provide simultaneous equations from which the electric charge of the third and fourth components and the value of the hypercharge $X$ of a quark multiplet can be obtained. In this way, for the model under consideration, the electric charge content of a quark $4$-plet is $(2/3, -1/3, 2/3, 5/3)$ and $X_q = 2/3$, and for a quark $\bar{4}$-plet we have $(-1/3, 2/3, -1/3, -4/3)$ and $X_{\bar{q}}=-1/3$.

Cancellation of the $[SU(3)_c]^3$ anomaly (or equivalently, the vectorlike character of $SU(3)_c$) requires the inclusion of the charge conjugate of each quark field as an $SU(4)_L \otimes U(1)_X$ singlet for which the hypercharge $X$ coincides with the electric charge. Cancellation of the $[SU(4)_L]^3$ anomaly requires to have equal number of $4$-plets and $\bar{4}$-plets, which in turn implies that we must have one family of quarks transforming under the representation $4$ and two families of quarks transforming under the representation $\bar{4}$ of $SU(4)_L$ so, since $X_{aL}=0$, the $X$ values obtained for the quark $4$-plets and $\bar{4}$-plets automatically ensure cancellation of the $[SU(4)_L]^2U(1)_X$ anomaly (provided $N_c=N_f$, where $N_c=3$ is the number of colors of $SU(3)_c$ and $N_f$ is the number of fermion families). Then, it is now just a matter of counting to check that the model is also free of the anomalies $[SU(3)_c]^2U(1)_X$, $[\mbox{grav}]^2U(1)_X$, and $[U(1)_X]^3$. The complete anomaly free fermion content of the model is then
\begin{eqnarray}\nonumber
L_{al}&=&(\nu_a, e_a, \nu^c_{a}, e^c_{a})_L\sim[1,4,0], \quad a=1,2,3;\nonumber \\
Q_{1L}&=&(u_1,d_1,U_1,J_1)_L\sim[3,4,2/3];\nonumber \\
u^c_{1L}&\sim&[\bar{3},1,-2/3];\quad d^c_{1L}\sim[\bar{3},1,1/3];\quad U^c_{1L}\sim[\bar{3},1,-2/3];\quad J^c_{1L}\sim[\bar{3},1,-5/3]; \nonumber \\
Q_{iL}&=&(d_i,u_i,D_i,J^\prime_i)_L\sim[3,\bar{4},-1/3],\quad i=2,3;
\nonumber \\ 
d^c_{iL}&\sim&[\bar{3},1,1/3];\quad u^c_{iL}\sim[\bar{3},1,-2/3];\quad D^c_{iL}\sim[\bar{3},1,1/3];\quad J^{\prime c}_{iL}\sim[\bar{3},1,4/3],\nonumber 
\end{eqnarray}
where $U$, $J$, $D_i$, $J^\prime_i$ are exotic quarks of electric charges $2/3$, $5/3$, $-1/3$, $-4/3$, respectively.

Masses for quarks and the symmetry breaking pattern $SU(3)_c\otimes SU(4)_L\otimes U(1)_X  \rightarrow  SU(3)_c\otimes SU(3)_L\otimes U(1)_X \rightarrow SU(3)_c\otimes SU(2)_L\otimes U(1)_Y  
\rightarrow SU(3)_c\otimes U(1)_Q$ are achieved by introducing the following four Higgs scalars \cite{su4b}: 
$\phi_1[1,4,-1]$ with a Vacuum Expectation Value (VEV) aligned in the direction $\langle\phi_1\rangle=(0,0,0,V^\prime)^T$; 
$\phi_2[1,4,0]$ with a VEV aligned as $\langle\phi_2\rangle=(0,0,V,0)^T$;
$\phi_3[1,4,1]$ with a VEV aligned in the direction $\langle\phi_3\rangle=(0,v^\prime,0,0)^T$, and
$\phi_4[1,4,0]$ with a VEV aligned as $\langle\phi_4\rangle=(v,0,0,0)^T$,
with the hierarchy $V\sim V^\prime >> v\sim v^\prime$.

To give masses to leptons, the Higgs decouplet $\phi_5[1,10,0]$ with the VEV
\[\left(\begin{array}{cccc}0 & 0 & 0 & 0\\
0 & 0 &  0 &  w \\
0 & 0 & 0 & 0\\
0 & w & 0 & 0 \end{array}\right),\]  
must be introduced. 

With $b=-1$ and $c=-4$, the 15 gauge fields associated with $SU(4)_L$ can be written as
\[\frac{1}{2}\lambda_\alpha A^\alpha_{\mu}=\frac{1}{\sqrt{2}}\left(
\begin{array}{cccc}D^{0}_{1\mu} & W^+_\mu & K^0_\mu & X^-_\mu\\ 
W^-_\mu & D^0_{2\mu} &  K^-_{1\mu} &  V^{--}_\mu\\
K^{\prime0}_\mu & K^+_{1\mu} & D^0_{3\mu} & Y^-_\mu\\
X^+_\mu & V^{++}_\mu & Y^+_\mu & D^0_{4\mu} \end{array}\right).\]

In view of the fact that the scalar $\phi_5$ is required only to give masses to leptons, $w$ can be assumed small as compared with $v$ and $v^\prime$ and its contribution to the gauge boson masses can be neglected. In this approximation, after breaking the symmetry with $\langle\phi_1\rangle +
\langle\phi_2\rangle+ \langle\phi_3\rangle+ \langle\phi_4\rangle$, the following masses for the charged gauge bosons are obtained:
$M^2_{W^\pm}=g^2(v^2+v^{\prime 2})/2$,
$M^2_{X^\pm}=g^2(v^2+V^{\prime 2})/2$, 
$M^2_{K^\pm_1}=g^2(v^{\prime 2}+V^2)/2$, 
$M^2_{Y^\pm}=g^2(V^2+V^{\prime 2})/2$,
$M^2_{K^0(K^{\prime 0})}=g^2(v^2 + V^2)/2$, 
and $M^2_{V^{++}(V^{--})}=g^2(v^{\prime 2}+V^{\prime 2})/2$. 
Since $W^\pm$ does not mix with the other charged bosons we have that 
$\sqrt{v^2+v^{\prime 2}}\approx 174$ GeV.

The mass matrix for the neutral gauge bosons, in the basis $(A_3^\mu, A_8^\mu, A_{15}^\mu, B^\mu)$, is given by \cite{su4b}

\[\frac{g^2}{2}\left(
\begin{array}{cccc} v^2+v^{\prime 2}& \frac{1}{\sqrt{3}}(v^2-v^{\prime 2})& \frac{1}{\sqrt{6}}(v^2-v^{\prime 2}) & -2 \delta v^{\prime 2}\\ 
\frac{1}{\sqrt{3}}(v^2-v^{\prime 2}) & \frac{1}{3}(v^2+v^{\prime 2}+4V^2) &  \frac{1}{3\sqrt{2}}(v^2+v^{\prime 2}-2V^2) &  \frac{2}{\sqrt{3}}\delta v^{\prime 2}\\
\frac{1}{\sqrt{6}}(v^2-v^{\prime 2}) & \frac{1}{3\sqrt{2}}(v^2+v^{\prime 2}-2V^2) & \frac{1}{3}(v^2+v^{\prime 2}+V^2+9V^{\prime 2}) & \frac{2}{\sqrt{6}}\delta (v^{\prime 2}+3V^{\prime 2})\\
-2 \delta v^{\prime 2} & \frac{2}{\sqrt{3}}\delta v^{\prime 2} & \frac{2}{\sqrt{6}}\delta (v^{\prime 2}+3V^{\prime 2}) & 4\delta^2 (v^{\prime 2}+V^{\prime 2})\end{array}\right),\]
where $\delta=g^\prime/g$. This mass matrix has one eigenvalue equal to zero corresponding to the photon. After extracting the photon, we are left over with a $3 \times 3$ mass matrix for three neutral gauge bosons
$Z^\mu$, $Z^{'\mu}$ and $Z^{''\mu}$ which, in the approximation 
$V^{\prime}= V >> v^{\prime}=v$, reads
\[g^2\left(
\begin{array}{ccc} \frac{1+4\delta^2}{1+3\delta^2}v^2& \sqrt{3}\delta^2\frac{\sqrt{1+4\delta^2}}{1+3\delta^2}v^2& 0 \\ 
\sqrt{3}\delta^2\frac{\sqrt{1+4\delta^2}}{1+3\delta^2}v^2 & \frac{2}{3}(1+3\delta^2)V^2 &  \frac{1}{3\sqrt{2}}\sqrt{1+3\delta^2}V^2 \\
0 & \frac{1}{3\sqrt{2}}\sqrt{1+3\delta^2}V^2 & 2 V^2 \end{array}\right),\]

The exact eigenvalues of this matrix are not very illuminating but, using the fact that $v/V << 1$, we can diagonalize it perturbatively. The standard procedure \cite{schiff} gives, at the first order in the perturbation parameter $q=(v/V)^2$, one eigenvalue of order $v^2$ associated to the neutral gauge boson of the SM and two eigenvalues of order $V^2$ associated to the two new neutral gauge bosons predicted by the model.

For this model, the photon field $A^\mu$ and the fields $Z_\mu$, $Z'_\mu$ and $Z^{''\mu}$ are given by
\begin{eqnarray} \nonumber
A^\mu&=&S_W A_3^\mu + C_W\left[\frac{T_W}{\sqrt{3}}\left(-A_8^\mu-
2\sqrt{2}A_{15}^\mu\right)+(1-3T_W^2)^{1/2}B^\mu\right]\;,\nonumber \\  
Z^\mu&=& C_W A_3^\mu - S_W\left[\frac{T_W}{\sqrt{3}}\left(-A_8^\mu-
2\sqrt{2}A_{15}^\mu\right)+(1-3T_W^2)^{1/2}B^\mu\right] \; , \nonumber \\ \label{fzzp3}
Z^{'\mu}&=&\frac{1}{3}(1-3T_W^2)^{1/2}\left(-A_8^\mu-
2\sqrt{2}A_{15}^\mu\right)-\sqrt{3}T_W B^\mu, \nonumber\\
Z^{''\mu}&=& \frac{2\sqrt{2}}{3}A^\mu_8-\frac{1}{3}A^\mu_{15},
\end{eqnarray}
where the sine of the electroweak mixing angle is $S_W = \delta/\sqrt{1+4\delta^2}$. We can also identify the $Y$ hypercharge associated with the SM abelian gauge boson as
\[ Y^\mu=\frac{T_W}{\sqrt{3}}\left(-A_8^\mu-
2\sqrt{2}A_{15}^\mu\right)+(1-3T_W^2)^{1/2}B^\mu.\]

\section{Conclusions}
In this paper we have done a detailed study of the $SU(3)_c\otimes SU(4)_L\otimes U(1)_X$ gauge symmetry. By restricting the gauge and fermion field representations to particles without exotic electric charges, eight different anomaly-free models have been identified. Four of them are three-family models in the sense that anomalies cancel by an interplay between the three families. Another two are one-family models where anomalies cancel family by family as in the SM. The remaining two are two-family models. Three-family models of this class have been recently proposed as viable models to implement the little Higgs mechanism \cite{little,kong}.

If we allow for particles with exotic electric charges, we end up with an infinite number of models. In this case, a particular model depends on the physical motivation of the 3-4-1 extension of the SM to be constructed. Examples of such a class of models are the ones studied in Refs.~\cite{su4a,su4b,cota,FaRi1,FaRi2}. In the main text we have presented the main features of the model developed in Ref.~\cite{su4b}. An interesting variation of this model, allowing the inclusion of the see-saw mechanism for neutrino masses and flavor mixing in the neutrino sector, is studied in Ref.~\cite{FaRi2}.

All the models presented here have in common two new neutral currents. One of them mixes with the SM neutral current which is also included as a part of each model. This mixing, however, is model dependent and has been calculated for models A and E of Sec.~\ref{sec3} in Refs.~\cite{pgs} and \cite{spp}, respectively, using experimental results from the CERN LEP, SLAC Linear Collider and atomic parity violation data. This calculation also shows that, for these two models, the mass of the new neutral gauge boson $Z_2^\mu$ which mixes with the one of the SM satisfies $0.7\;\mbox{TeV}\leq M_{Z_2}$, which is compatible with the bound obtained from $p\bar{p}$ collisions at the Fermilab Tevatron.

It is worth noticing that both for the particular 3-4-1 model with exotic electric charges presented in Sec.~\ref{sec4}, and for all the four three-family models without exotic electric charges (models A, B, E and F in Sec.~\ref{sec3}), universality for the known leptons in the three families is present at the tree level in the weak basis, up to mixing with the exotic fields. Since the mass scale of the new neutral gauge boson $Z^{\prime}$ and of the exotic particles is of the order of $V$ (the scale of the $SU(4)_L\otimes U(1)_X$ symmetry), this mixing will suppress tree-level flavor changing neutral currents (FCNC) effects in the lepton sector. For the quarks, instead, one family transform differently from the other two and, as a result, there can be potentially large FCNC effects in the hadronic sector.

Finally, let us mention that all the 3-4-1 models without exotic electric charges identified in this work are relatively new in the literature. In particular, the phenomenology of models B, D, F, H, I, has not yet been studied, as far as we know.

\section*{Acknowledgments}

Work partially supported by Universidad de Antioquia, and by Universidad Nacional de Colombia-Sede Medell\'\i n.

\end{document}